\documentclass[11pt]{article}
\usepackage{amsmath,amsthm,latexsym,amssymb,amsfonts,parskip,epsfig,xspace}
\usepackage[a4paper,top=2cm,bottom=3cm]{geometry}

\usepackage{palatino}
\usepackage[mathcal]{euler}

\newtheorem{theo}{Theorem}

\newcommand{\R}{\mathbb{R}}                  
\newcommand{\C}{\mathbb{C}}                  

\newcommand{\scpr}[2]{\langle#1\, \vert \, #2 \rangle}

\newcommand{\betr}[1]{\left\lvert #1 \right\rvert}

\newcommand{\expec}[1]{\langle #1 \rangle}

\newcommand{\hilb}[1]{\mathcal{#1}}

\newcommand{\der}[2]{\frac{\text{d} #1}{\text{d} #2}}

\newcommand{\SLTC}{SL(2,$\C$)}
\newcommand{\SUT}{SU(2)\xspace}
\newcommand{\sut}{su(2)\xspace}

\newcommand{\lqg}{loop quantum gravity\xspace}
\newcommand{\Lqg}{Loop quantum gravity\xspace}

\newcommand{\lqc}{loop quantum cosmology\xspace}
\newcommand{\hdiff}{\hilb{H}_\text{diff}}
\newcommand{\hkin}{\hilb{H}_\text{kin}}
\newcommand{\hphys}{\hilb{H}_\text{phys}}
\newcommand{\hgauge}{\hilb{H}_\text{gauge}}
\newcommand{\diff}{\text{Diff}}
\newcommand{\tdiff}{\text{TDiff}}

\DeclareMathOperator{\cyl}{\text{cyl}}
\DeclareMathOperator{\tr}{\text{Tr}}

\newcommand{\one}{\mathbb{I}}

\numberwithin{equation}{section}
\allowdisplaybreaks

\title{Loop Quantum Gravity -- a short review\footnote{Talk delivered at the workshop
``Foundations of Space and Time -- Reflections on Quantum Gravity'' in honor of
George Ellis, STIAS, Stellenbosch, South Africa, 10-14 August 2009.}}
\author{Hanno Sahlmann\\[.5cm] Institute for Theoretical Physics, Karlsruhe
University\\
Karlsruhe Institute for Technology\\
}
\date{{\small Preprint KA-TP-19-2009}}
\begin{document}
\maketitle
\begin{abstract}
In this article we review the foundations and the present status of \lqg. 
It is short and relatively non-technical, the emphasis is on the ideas, and  
the flavor of the techniques. In particular, we describe the kinematical
quantization and the implementation of the Hamilton constraint, as well as the
quantum theory of black hole horizons, semiclassical states, and matter
propagation. Spin foam models and loop quantum cosmology are mentioned only in
passing, as these will be covered in separate reviews to be published alongside
this one. 
\end{abstract}
\section{Introduction}
\Lqg is non-perturbative approach to the quantum theory of gravity, in which no
classical background metric is used. In particular, its starting point is 
not a linearized theory of gravity. As a consequence, while it still operates
according to the rules of quantum field theory, the details are quite different
of those for field theories that operate on a fixed classical background
space-time. It has considerable successes to its credit, perhaps most notably a
quantum theory of spatial geometry in which quantities such as 
area and volume are quantized in units of the Planck length, and a calculation
of black hole entropy for static and rotating, charged and neutral black holes.
But there are also open questions, many of them surrounding the dynamics
(``quantum Einstein equations'') of the theory.  

In contrast to other approaches such as string theory, \lqg is rather modest in
its aims. It is not attempting a grand unification, and hence is not based on an
overarching symmetry principle, or some deep reformulation of the rules of
quantum field theory. Rather, the goal is to quantize Einstein gravity in four
dimensions. While, as we will explain, a certain amount of unification of the
description of matter and gravity is achieved, In fact, the question of whether
matter fields must have special properties to be consistently coupled to gravity
in the framework of \lqg is one of the important open questions in \lqg. 

\Lqg is, in its original version, a canonical approach to quantum gravity.
Nowadays, a covariant formulation of the theory exists in the so called
\emph{spin foam models}. One of the canonical variables in \lqg is a connection,
and many distinct technical features (such as the `loops' in its name) are
directly related to the choice of these variables. Another distinct feature of
\lqg is that no fixed 
classical geometric structures are used in the construction. New techniques had
to be developed for this, and the resulting Hilbert spaces that look very
different than those in standard quantum field theory, with excitations of the
fields one- or two-dimensional. But it has also simplified the theory, since can
be shown that some choices made in the quantum theory are actually uniquely
fixed by the requirement of background independence. Furthermore, the
requirement of background independence seems to lead to a theory which is built
around a very quantum mechanical gravitational ``vacuum'', a state with
degenerate and highly fluctuating geometry. This is exciting, because it means
that when working in \lqg, the deep quantum regime of gravity is `at one's
fingertips'. However, it also means that to make contact with low energy physics
is a complicated endeavor. The latter problem has attracted a considerable
amount of work, but is still not completely solved. Another (related) challenge
is to fully understand the implementation of the dynamics. 
In \lqg the question of finding quantum states that satisfy `quantum Einstein
equations' is reformulated as  finding states that are annihilated by the
quantum Hamilton constraint. The choices that go into the definition of this
constraint are poorly understood in physical terms. Moreover the constraint
should be implemented in an anomaly-free way, but what this entails in practice,
and whether existing proposals fulfill this requirement are still under debate.
This is partially due to the lack of physical observables with manageable
quantum counterpart, to test the physical implication of the theory. 
  
While these challenges remain, remarkable progress has happened over the last
couple of years: The master constraint program has brought new ideas to bear on
the implementation of the dynamics \cite{Thiemann:2003zv}. Progress has been made in identifying observables for general relativity that can be used in the canonical quantization \cite{Dittrich:2005kc,Dittrich:2006ee,Giesel:2007wi,Domagala:2010bm}. A revision of the  vertex amplitudes used in spin foam
models has brought them in much more direct contact to \lqg \cite{Engle:2007uq,Engle:2007wy,Kaminski:2009fm,Kaminski:2009cc}. And, last
not least, in \lqc, the application of the quantization strategy of \lqg to
mini-superspace models has become a beautiful and productive laboratory for the
ideas of the full theory, in which the quantization program of \lqg can be
tested, and, in many cases, brought to completion  \cite{Bojowald:2008zzb,Bojowald:2008gz,Ashtekar:2006rx,Ashtekar:2009mb,Kaminski:2008td}. The present review
will not cover these developments in any detail, partially because they are
ongoing, and partially because there will be separate reviews on group field
theory and \lqc published alongside the present text. But we hope that it makes
for good preparatory reading. In fact, the basic connection between \lqg and
spin foam models is explained in section \ref{se_sfm}, the master constraint
program is briefly described in section  \ref{se_mc}, and there are some
references to \lqc in section \ref{se_app}. Certainly the present review can
also not replace the much more complete and detailed reviews that are available.
We refer the interested reader in particular to \cite{Ashtekar:2004eh,Thiemann:2007zz, Rovelli:2008zza}. 

The structure of the review is as follows: In section \ref{se_kin} we explain
the classical theory and kinematical quantization underlying \lqg. Section
\ref{se_ham} covers the implementation of the Hamilton constraint. In section
\ref{se_app} we consider some physical aspects of the theory: quantized black
hole horizons, semiclassical states, and matter propagation. We close with an
outlook on open problems and new ideas in section \ref{se_out}.   

\section{Kinematical setup}
\label{se_kin}
\Lqg is a canonical quantization-approach to general relativity, thus it is
based on a splitting of space-time into time and space, and on a choice of
canonical variables. Implicit in the splitting is the assumption that the
space-time is globally hyperbolic. Whether topology change can nevertheless be
described in the resulting quantum theory is a matter of debate. The choice of
canonical variables is characteristic to \lqg: One of the variables is a
connection, and hence the phase space (before implementation of the dynamics)
has the same form as that of Yang-Mills theory. As with any canonical
formulation of general relativity, the theory has constraints that have to be
handled properly both in the classical and in the quantum theory. 

The quantization strategy applied in \lqg is that of Dirac, for the case of
first class constraints: First, a \emph{kinematical} representation of the basic
fields by operators on a Hilbert space $\hilb{H}_\text{kin}$ is constructed. In
this representation, operators corresponding to the constraints are defined.
Then, quantum solutions to the constraints are sought. Such solutions, also
called \emph{physical} states, are quantum states that are in the kernel of all
the constraints. They form the physical Hilbert space $\hilb{H}_\text{phys}$.
Finally, observable quantities are quantized. The corresponding operators should
form an algebra $\mathfrak{A}$, and commute with the quantum constraints. Thus
$\mathfrak{A}$ leaves  $\hilb{H}_\text{phys}$ invariant. The pair
$(\mathfrak{A}, \hilb{H}_\text{phys})$ then constitutes the quantum theory of
the constrained system in question.
Technical aspects of this procedure have to be refined in \lqg. For example, if
the zero eigenvalue in the continuous part of the spectrum of one of the
constraints, the resulting physical space is not part of the Hilbert space but
part of its dual. But there are also some fundamental questions about this
procedure, such as what guides the choice of the kinematical Hilbert space, and
how the quantization and implementation of the constraints is checked. Also, it
is notoriously difficult to write down explicit examples of observables for
general relativity in the canonical setting, even in the classical theory.  

While some of the above questions are not yet answered for \lqg, the quantum
theory is successful in many respects: It includes a fully quantized  spatial
geometry, and an implementation of the constraints that is anomaly-free at least
in a certain sense. In the following, we will give a short, and mostly
non-technical introduction to the kinematical aspects of the quantization. The
quantization of the Hamilton constraint will be discussed in section
\ref{se_ham}.
\subsection{Connection formulation of general relativity}
\Lqg rests on a reformulation of ADM canonical gravity in terms
of variables similar to 
those of Yang-Mills theory. Ashtekar discovered a formulation
\cite{Ashtekar:1986yd} in terms of a self-dual \SLTC connection, and its
canonical conjugate, satisfying suitable reality conditions. \Lqg came to use a
formulation in terms of an \SUT connection \cite{Barbero:1994ap} for technical
reasons. Both of these are actually special cases of a family of formulations
depending on several parameters (\cite{Rezende:2009sv} and literature given
there). We will only consider one of these, the Barbero-Immirzi parameter
$\iota$ \cite{Immirzi:1996di}. The covariant description in this case is the
Holst-Action
\begin{equation}
\label{eq_holst}
S[e,\omega]=\int\epsilon^{IJKL}e_I\wedge e_J \wedge
F_{IJ}(\omega)+\frac{1}{\iota} e^{I}\wedge e^J \wedge F_{IJ}(\omega)
\end{equation}
for an \SLTC connection $\omega$ and a vierbein $e$. In the limit $\iota
\rightarrow \infty$, this is the well known Palatini action of general
relativity. The so called \emph{Holst term} proportional to $\iota^{-1}$ is not
a topological term, it depends on the geometry. But, in the absence of fermionic
matter, it does not change the equations of motion, as it vanishes identically
on shell, due to the Bianchi identity.\footnote{Actually, instead of adding this
term, one can also add the Nieh-Yang term, which \emph{is} topological. The
resulting canonical formulation is the same as that with a real Barbero-Immirzi
parameter.} In the presence of fermions, there are small effects that could in
principle be used to distinguish the formulation \eqref{eq_holst} from the
Palatini formulation \cite{Perez:2005pm}.

The Holst-term has a profound effect on the canonical formulation of the theory.
A Legendre transform of the Palatini action leads (after solving the
second-class 
constraints) back to the 
ADM-formulation, with spatial metric and exterior curvature as canonical
variables. The Legendre-transform of \eqref{eq_holst} with \emph{finite}  
Barbero-Immirzi parameter leads, however, to formulations in which one canonical
variable is a connection: For $\iota=\pm i$ the theory has special symmetries
and one obtains the Ashtekar formulation \cite{Ashtekar:1986yd} in terms of a
self-dual \SLTC connection. For real $\iota$, and after a partial gauge fixing
that gets rid of second class constraints, one obtains a canonical pair
consisting of an \SUT connection $A_a^I$ and a corresponding canonical momentum
$E_J^b$, 
\begin{equation}
\label{eq_poi}
 \{A_{a}^I(x),E^b_J(y)\}=8\pi G \iota\, \delta_a^b\delta^I_J\delta(x,y). 
\end{equation}
These fields take values on a spatial slice $\Sigma$ of the manifold that was
chosen in the process of going over to the Hamilton formulation. 

There are several constraints on these variables, and the Hamiltonian is a
linear combination of constraints. The equations for time evolution are the
usual Hamilton equations, and together with the constraint equations they form
a set of equations which is completely equivalent to Einstein's equations. 
The constraints can be written in the following way: 
\begin{align}
 \label{eq_const}
 G_I&=D_aE^a_I\\
C_a&=E^b_IF^I_{ab}\\
H&=\frac{1}{2}\epsilon^{IJ}{}_K\frac{E^a_I E^b_J}{\sqrt{\det E}}F^K_{ab}
-(1+\iota^2)\frac{E^a_I E^b_J}{\sqrt{\det E}}K_{[a}^IK_{b]}^J
\end{align}
where $D$ is the covariant derivative induced by $A$, $F$ is the curvature of
$A$, and $K$ is the extrinsic curvature of $\Sigma$ in space-time.  
They have a simple geometric interpretation: $G_I$ generates gauge
transformations on phase space. It is also called \emph{Gauss constraint} to
highlight that it is completely analogous to the Gauss-law constraint that
shows up in electrodynamics. $C_a$ generates the transformations induced in
phase space under diffeomorphisms of $\Sigma$. It is therefore also called
\emph{diffeomorphism constraint}. Finally, $H$ generates (when the other
constraints hold) the transformations induced in phase space under
deformations of (the embedding of) the hypersurface $\Sigma$ in a timelike
direction in space-time. It is also called the \emph{Hamiltonian constraint},
since such deformations can be interpreted as time evolution.  

The canonical momentum $E$ has a direct geometric interpretation: It encodes
the spatial geometry:
\begin{equation}
  |\det q| q^{ab}=E^a_IE^b_J \delta^{IJ}
\end{equation}
where $q_{ab}$ is the metric induced on $\Sigma$ by the space-time metric. 
Thus $E$ is a densitized triad field for $q$. 
The interpretation of $A$ is slightly more involved. 
\begin{equation}
A_a^I=\Gamma_a^I+\iota K_a^I 
\end{equation}
where $\Gamma$ is the spin connection related to $E$. 

Matter fields can be added to the canonical description  given above. This has
to be done with some care, so as to not change the structure of the
gravitational
sector. For the fermionic sector this requires working with slightly unusual
(``half density'') variables \cite{Thiemann:1997rq}.
\subsection{Kinematic representation}
The basic variables for the quantization in \lqg are chosen in such a way as to
make their transformation behavior under \SUT and spatial diffeomorphisms as
simple and transparent as possible. The obvious reason behind this goal is that
one wants to simplify the solution of the constraints as much as possible. 
Important early ideas about this are in Gambini, Trias, Rovelli and Smolin \cite{Gambini:1986ew, Rovelli:1989za}. The rigourous implementation of the program is developed in 
\cite{Ashtekar:1991kc,Ashtekar:1993wf,Lewandowski:1993pc,Ashtekar:1994mh,Ashtekar:1994wa}. An
obvious choice for the connection $A$ are its holonomies
\begin{equation}
h_{\alpha}[A]=\mathcal{P} \exp \int_{\alpha} A, 
\end{equation}
or more generally, functions of such holonomies, 
\begin{equation}
f[A]\equiv f(h_{\alpha_1}[A],h_{\alpha_2}[A],\ldots,h_{\alpha_n}[A])
\end{equation}
for a finite number of paths $\alpha_1,\ldots, \alpha_n$. Such functionals are
also called \emph{cylindrical functions}. 

For the field $E$ a natural functional is its flux through surfaces $S$ \cite{Ashtekar:1996eg}:
\begin{equation}
\label{eq_flux}
E_[S,f]=\int_S *E_I f^I
\end{equation}
where $f$ is a function taking values in \sut$^*$ and *E is the two-form 
$E^a \epsilon_{abc}\text{d}x^b \wedge \text{d}x^c$. 

To quantize cylindrical functions and fluxes, one is seeking a representation of
the following algebraic relations on a Hilbert space: 
\begin{equation}
\label{eq_qalg}
\begin{split}
 f_1\cdot f_2[A]&=f_1[A]f_2[A]\\
  [f,E_{S,r}]&= 8\pi \iota l^2_P X_{S,r}[f]\\
  [f,[E_{S_1,r_1}, E_{S_2,r_2}]]&=(8\pi \iota l^2_P)^2[X_{S_1,r_1},
X_{S_2,r_2}][f]  \\
  \ldots &\\
  (E_{S,r})^*=E_{S,\overline{r}},\quad&\quad
  (f[A])^*=\overline{f}[A]
  \end{split}
\end{equation}
Here, $X$ is a certain derivation on the space of cylindrical functions \cite{Ashtekar:1996eg}. As an
example, consider the case of a surface $S$ that is intersected transversally by
a path $e$, splitting it into a part $e_1$ incoming to, and a part $e_2$
outgoing from the surface. Then (with a certain orientation of the surface
assumed)    
\begin{equation}
\label{eq_der}
X_{S,r} \overset{j}\pi(h_{e})=\sum_i r_i(p) \overset{j}{\pi}(h_{e_1}{
\tau_i}h_{e_2}).
\end{equation}
\begin{figure}%
\centerline{\includegraphics[width=8cm]{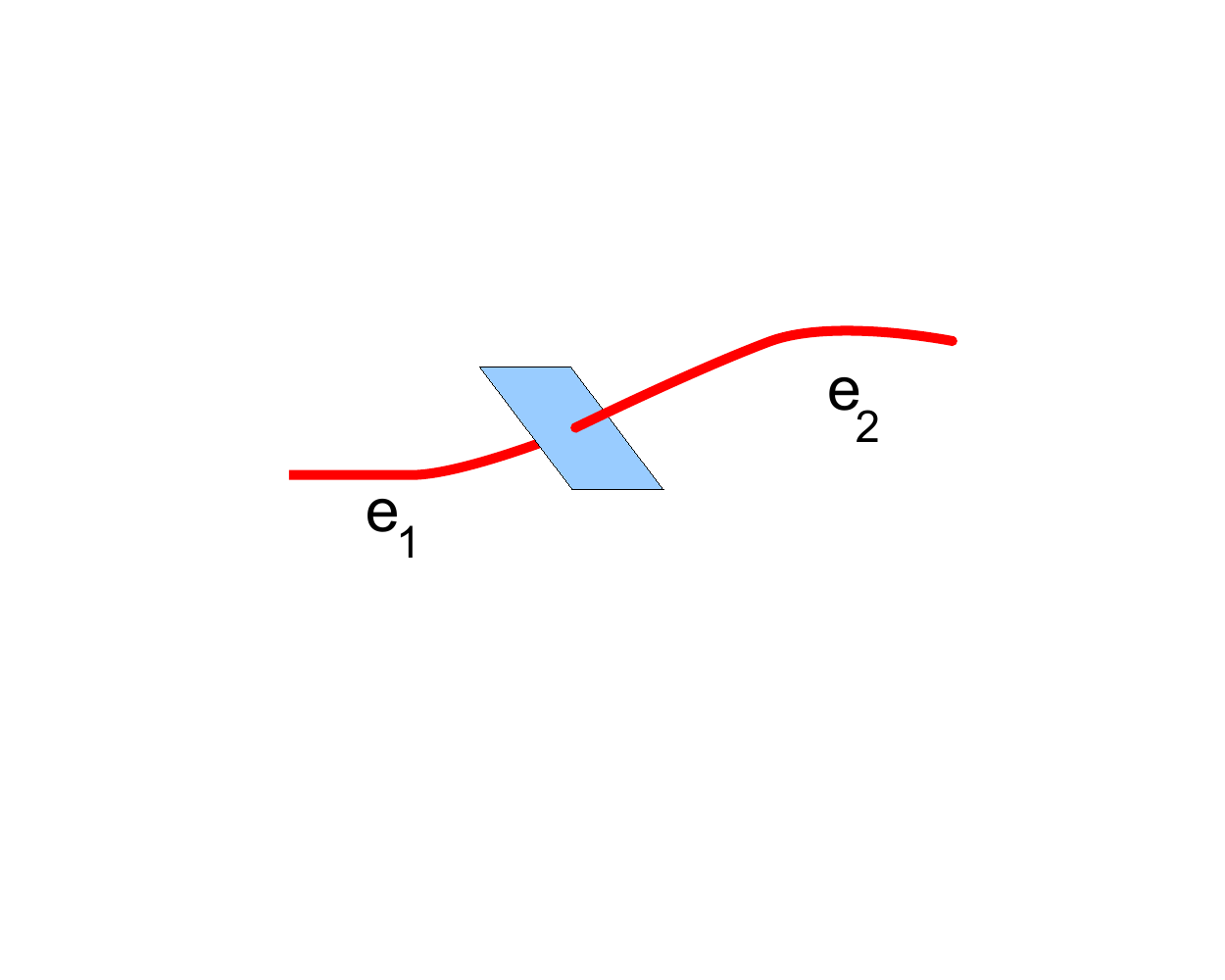}}%
\caption{Illustration of the edges involved in \eqref{eq_der}}%
\label{fi_der}%
\end{figure}
The commutators between cylindrical functions and fluxes come from the Poisson
relations \ref{eq_poi}. 
It is somewhat surprising to see that there are also non-trivial commutators
between fluxes. These are required to turn the algebra of fluxes and cylindrical
functions into a Lie-algebra, a structure that has representations in terms of
operators on Hilbert-spaces \cite{Ashtekar:1998ak}.  

\begin{figure}%
\begin{center}
\includegraphics[width=9cm]{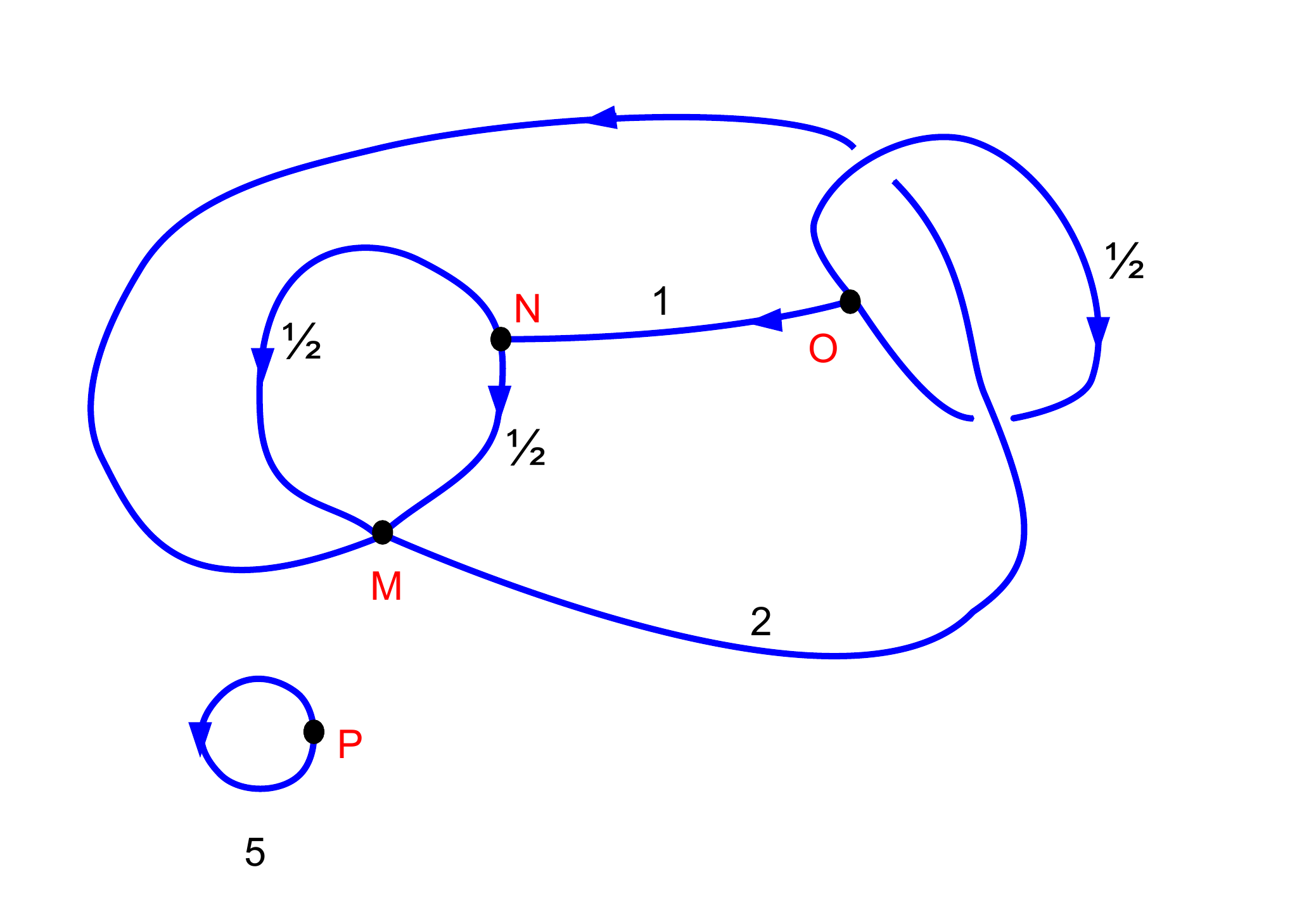}%
\end{center}
\caption{A generalized spin network. The labels M,N,O,P can be thought of as
suitably dimensional tensors.}%
\label{fi_gsn}%
\end{figure}

\Lqg employs a specific representation of \eqref{eq_qalg} on a Hilbert space
$\hilb{H}_\text{kin}$. A basis for this Hilbert space is given by the so called
generalized spin networks. Such a network is by definition an oriented graph
$\gamma$ embedded in $\Sigma$, together with a labeling of the edges and
vertices of that graph: The edges are labeled by irreducible representations of
$\SUT$. A vertex carries elements of the dual of the tensor product of all
representations on the edges that are incoming to or outgoing from the vertex as
a label (see figure \ref{fi_gsn}).  
A generalized spin network represents a way of constructing a cylindrical
functional. To obtain its value on a given connection, one computes the
holonomies along the edges of the graph in the representations given by the edge
labels, and contracts these via the labels of the vertices. And, vice versa, any cylindrical function can be written as a (possibly infinite) linear combination
of generalized spin networks.  

An inner product is then defined on the span of these generalized spin networks
by postulating that they are orthogonal to each other, and by specifying their
norm in terms of the labels of the graph. This inner product is completely
invariant under the action of the diffeomorphisms of $\Sigma$. 

A representation of \eqref{eq_qalg} is given on generalized spin
networks by using the fact that they can be viewed as cylindrical functionals.
The cylindrical functionals can thus be represented as multiplication
operators, the fluxes by derivations 
\begin{equation}
 (f\psi)[A]=f[A]\psi[A], \qquad (E_{S,r}\psi)[A] = (X_{S,r}\psi)[A]. 
\end{equation}
As with the inner product, these definitions do not make use of any background
structure, such as a classical metric. They are thus covariant under the action
of the diffeomorphisms of $\Sigma$. Moreover, there is a state in $\hkin$ that
is even \emph{invariant} under the action of those diffeomorphisms. This state
is the empty (i.e. without any edges or vertices) generalized spin network.
Moreover, any state in the kinematical Hilbert space can be approximated by
applying linear combinations of products of the basic operators to this
diffeomorphism invariant state. In mathematical language, this state is
therefore cyclic. It, together with the algebraic relations between the basic
variables completely encodes the structure of the kinematical representation. 

We note also that the kinematical representation has the following peculiar
properties:
\begin{enumerate}
	\item The diffeomorphisms $\phi$ of $\Sigma$ are represented on $\hkin$
by unitary operators $U_\phi$. This follows from what we have already said about
their action. But generators for these unitary operators do not exist. If
$\phi(t)$ is a one parameter family of diffeomorphisms, with $\phi(0)=\one$,
then 
\begin{equation}
\label{eq_nogen}
   \frac{1}{i}\left.\der{}{t}\right\rvert_0 U_{\phi(t)}
\end{equation}  
does not exist, in any sense, as a well defined operator. 
  \item We have seen that the holonomies $h_e[A]$ exist as matrices of
operators. But neither can one obtain from them an operator for the curvature
$F$, nor for the connection $A$ itself: The limits
\begin{equation}
\label{eq_shrink}
\lim_{\epsilon\rightarrow 0} 
\frac{1}{\epsilon^2}\left(h_{\alpha_\epsilon}-\one\right),\qquad
 \lim_{\epsilon\rightarrow 0}
\frac{1}{\epsilon}\left(h_{\beta_\epsilon}-\one\right)
\end{equation}
\begin{figure}%
\centerline{\includegraphics[width=8cm]{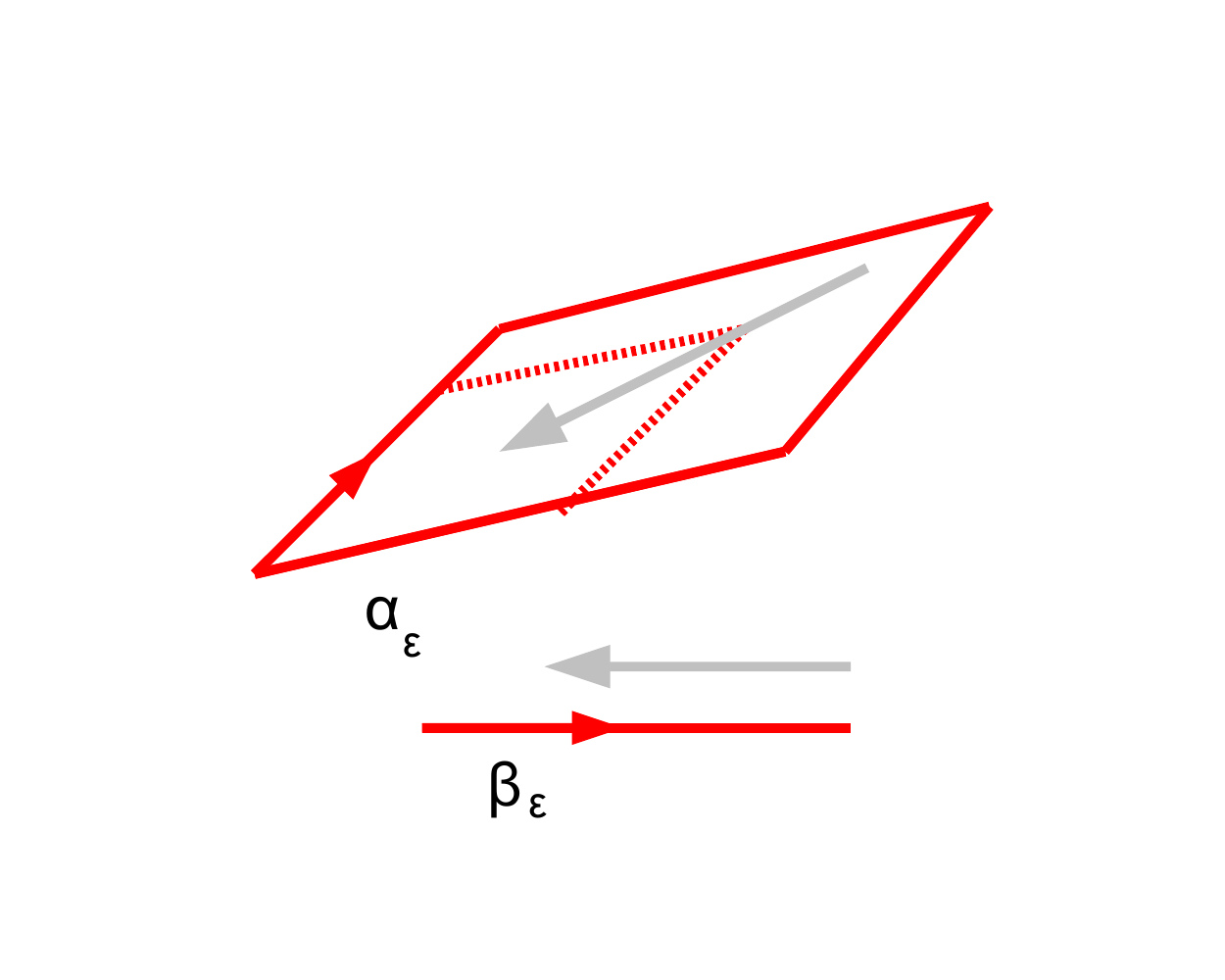}}%
\caption{The loop $\alpha_\epsilon$ and the edge $\beta_\epsilon$ of \eqref{eq_shrink}}%
\label{fi_shrink}%
\end{figure} 
do not exist in any sense as well defined operators on $\hkin$.
$\alpha_\epsilon$ is here a plaquette loop with (coordinate) side length 
$\epsilon$, and $\beta_\epsilon$ is an open line with (coordinate) side length 
$\epsilon$ (see figure \ref{fi_shrink}).
\end{enumerate}
It may appear that a lot of choices have been made in the definition of $\hkin$
and the representation of the basic variables on it. But this is not the case.
The following uniqueness theorem can be proven \cite{Lewandowski:2005jk,Fleischhack:2004jc}. 
\begin{theo}
Any representation of the algebraic relations \eqref{eq_qalg} that contains a
diffeomorphism invariant cyclic vector is equivalent to the one on $\hkin$
described above.  
\end{theo}
Diffeomorphism invariance should be seen here as a requirement dictated by the
philosophy of \lqg (no use of geometric background structure), as well as by
simplicity (implementation of the diffeomorphism constraint consists precisely
in throwing out any non-diffeomorphism invariant information). While cyclicity
would be a requirement on the physical sector, here it is only a natural
simplification.

\subsection{Geometric operators}
It is possible to quantize areas and volumes with respect to the geometry on
$\Sigma$ on the Hilbert space $\hkin$ 
\cite{Rovelli:1994ge,Ashtekar:1996eg,Lewandowski:1996gk,Ashtekar:1997fb}. Since the quantum Einstein equations, in
the form of the constraints, have not yet been taken into account, the
physical implications of the results have to be considered with substantial
care \cite{Dittrich:2007th,Rovelli:2007ep}. There are, however situations, in which such quantities are observables,
in the sense that they commute with the constraints. This is for example the
case with the area of a black hole horizon as considered in section \ref{se_bh}
below. In such cases the results that we are going to present have clear
physical significance. 

We consider the case of areas: Let $S$ be a surface in $\Sigma$. When the field
$E$ is pulled back to $\Sigma$ one obtains a vector valued two-form. The norm
of this two-form is directly related to the area \cite{Rovelli:1993vu}:
\begin{equation}
A_S=\int_S |E|.   
\end{equation}
This formula can be used as a starting point for quantization. Regularizing in
terms of fluxes in the form of \eqref{eq_flux}, substituting operators, and
taking the regulator away leads to a well defined, simple operator
$\widehat{A}_S$. Its action on states with just a single edge is especially 
simple: If edge and surface do not intersect, the state is annihilated. If they
do intersect once, one obtains 
\begin{equation}
\label{eq_aspec}
\widehat{A}_S\tr[\pi_j(h_\alpha[A])]=8\pi\iota {
l_P^2}\sqrt{{j(j+1)}}\tr[\pi_j({h}_\alpha[A])].
\end{equation}
Thus these states are eigenstates of area, with the eigenvalue given as the
square root of the eigenvalue of the \SUT-Casimir in the representation given
on the edge. A slightly more complicated action is obtained in the case of
several intersections, and in particular if a vertex of the generalized spin
network lies within the surface. Nevertheless the area operator can be
completely diagonalized. It turns out that the spectrum is 
discrete. As is seen in \eqref{eq_aspec}, the scale is set by Planck area
$l_P^2$. The eigenvalue-density increases exponentially with area. 
\begin{figure}%
\centerline{\includegraphics[width=8cm]{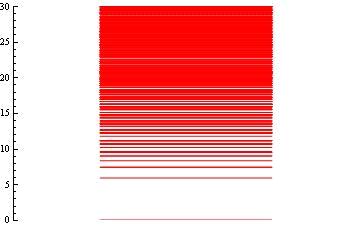}}%
\caption{The lowest part of the area spectrum of \lqg, in units of $l_P^2$ }%
\label{fi_areaspec}%
\end{figure}
A similar procedure leads to an operator for volumes of subregions in $\Sigma$.
This operator is substantially more complicated. Unlike the area operator,
the action of which is purely in terms of the representation label of the
edges, the volume operator acts on the vertices, by changing the maps that
label them (``recoupling''). In fact, there are two slightly
different versions of the volume operator \cite{Rovelli:1994ge,Ashtekar:1997fb}, differing in the way the
tangent space structure of a vertex is taken into account. In either case, the
spectrum is discrete. but not explicitly known. Some remarkable analytic developments are in \cite{Thiemann:1996au,Brunnemann:2004xi,Brunnemann:2010yv}. A beautiful computer analysis
of the lowest part of the spectrum can be found in \cite{Brunnemann:2006kx,Brunnemann:2007ca}. 

\subsection{Gauge invariant states, spin networks}
The simplest of the constraints \eqref{eq_const} to implement is the Gauss
constraint. $G_I=D_aE^a_I$ it can be easily checked that classically, it
generates \SUT transformations, which act on holonomies as
\begin{equation}
\label{eq_gt}
h_e[A]\mapsto g(s(e)) h_e g(t(e))^{-1} 
\end{equation}
with $g(x)$ the gauge transformation, and $s(e),t(e)$ the beginning and
endpoint of $e$. Thus, there are two ways to implement this
constraint: One can regularize the expression for $G_I$ in terms of holonomies
and fluxes, which have well defined quantization, quantize the regularized
expression, and remove the regulator, hoping to obtain a well defined
constraint operator in the limit. If successful, one can then determine the
kernel of the quantum constraint. Or one can declare that all states in $\hkin$
that are invariant under gauge transformations \eqref{eq_gt}, are
solutions to the constraint. Both strategies are viable, and lead to exactly
the same result: The solution space $\hgauge$ is a proper subspace of $\hkin$.
An orthonormal basis is given by the so called \emph{spin networks} \cite{Baez:1994hx,Rovelli:1995ac}. These are 
special cases of the generalized spin networks, in that the linear maps
labeling the vertices are intertwining operators 
\begin{equation}
I_v: \bigotimes_{e \text{ incoming}} \pi_{j(e)} \longrightarrow \bigotimes_{e
\text{ outgoing}} \pi_{j(e)}, \qquad I_v\pi_{\text{ incoming}}(g)=
 \pi_{\text{ outgoing}}(g)I_v
\end{equation}
mapping the tensor product of the representations on the incoming edges to the
tensor product of the representations on the outgoing edges. The contraction
of the holonomies with these intertwiners guarantees that the resulting states
are invariant under gauge transformations.
\subsection{Diffeomorphism invariant states}
The diffeomorphism constraint $C_a=E^b_IF^I_{ab}$ has not been quantized
directly. One reason is that curvature can not be quantized on $\hkin$ but one
can see even on more general grounds that a quantization of $C_a$ must run into
difficulties: Classically, this constraint generates the diffeomorphisms of
$\Sigma$, and one expects the same of its quantum counterpart. Otherwise one
would have produced an anomalous implementation of the constraint, with
possibly disastrous consequences for the theory. But the diffeomorphisms
$\phi$ of $\Sigma$ already act on $\hkin$, through unitary operators $U_\phi$.
These operators are however, not strongly continuous in the diffeomorphisms
(see \eqref{eq_nogen}), in other words, they have no selfadjoint generators.  
Thus $C_a$ can not be directly quantized without generating anomalies. But this
is not a problem, as we know what the gauge transformations generated by  $C_a$
are, and because they are acting in a simple manner on $\hkin$. All one has to
do is find states that are invariant under the action of the diffeomorphisms
$U_\phi$. 

The action of the diffeomorphisms on cylindrical functions consists in moving
the underlying graph: 
\begin{equation}
 U_\varphi\psi_{\gamma}=\psi_{\varphi(\gamma)}
\end{equation}
Therefore, the only invariant state in $\hgauge$ is the empty spin network.
Rather than in $\hdiff$, the rest of the invariant states in lying in the dual
of $\hdiff$. They can be found by group averaging. This procedure assigns to a 
state $\psi\in \hgauge$ a diffeomorphism invariant functional $\Gamma\psi$. The
idea is  
\begin{equation}
(\Gamma \psi)(\phi) = (\text{Vol}(\diff))^{-1} \int_{\diff}
D\varphi\, \scpr{U_\varphi\psi}{\phi}_{\hkin}.
\end{equation}
This is still formal. To make this work, the integration over the
diffeomorphism group, and the division by its volume, have to be made sense of.
These tasks would be hopeless, were it not for the unusual properties of the 
scalar product on $\hkin$. In fact, the correct notion in this context of the
integral over diffeomorphisms is that of a sum! A careful examination leads to
the formula  \cite{Ashtekar:1995zh, Ashtekar:2004eh}
\begin{equation}
\label{eq_rig}
 (\Gamma \psi_\gamma)(\phi)=\sum_{\varphi_1\in
\diff/\diff_\gamma}\frac{1}{\betr{\text{GS}_\gamma}}
 \sum_{\varphi_2\in \text{GS}_\gamma}\scpr{\varphi_1\ast\varphi_2\ast
\psi_\gamma}{\phi}.
\end{equation}
Here,  $\diff_{\gamma}$ is the
subgroup of diffeomorphisms mapping $\gamma$ onto itself, and $\tdiff_\gamma$
the subgroup of $\diff$ which is the identity on $\gamma$. The quotient
$\text{GS}_\gamma :=\diff_\gamma / \tdiff_\gamma$ is called the set of
\emph{graph symmetries}. It can be checked that this definition really
gives diffeomorphism invariant functionals over $\hgauge$. An inner product can
also be defined on these functionals, using \eqref{eq_rig}. Thus one obtains a
Hilbert space $\hdiff$ of gauge and diffeomorphism invariant quantum states.  

\begin{figure}%
\centerline{\includegraphics[width=8cm]{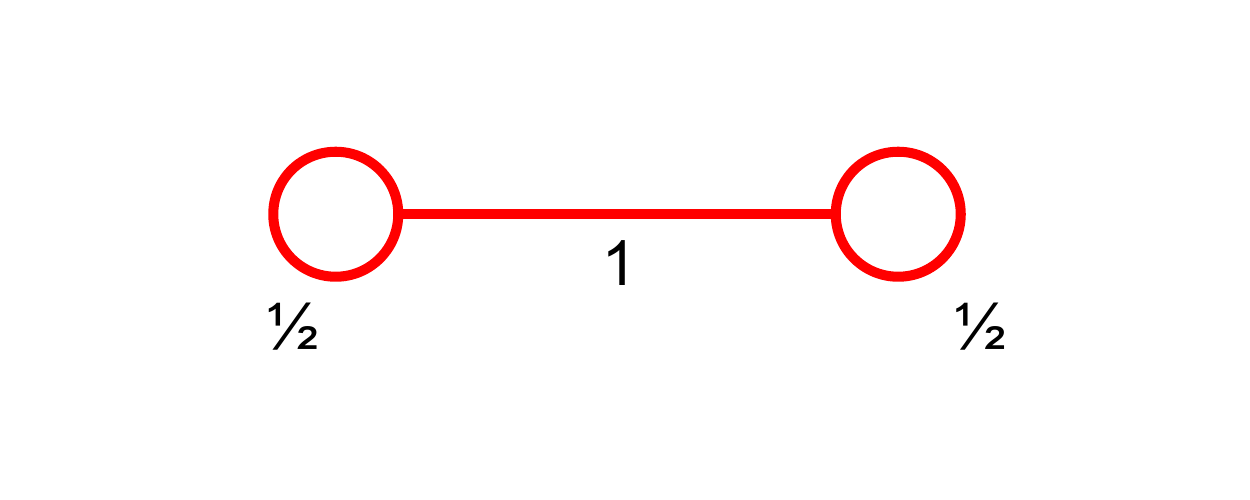}}%
\caption{The hourglass spin network gets mapped to zero under group averaging with respect to the diffeomorphism group}%
\label{fi_hour}%
\end{figure}

It is sometimes stated that diffeomorphism invariant spin network states are
labeled by equivalence classes of spin networks under diffeomorphisms. This is
a nice intuitive picture, but one has to be careful with it: The effects of
\eqref{eq_rig} can be quite subtle. For example, the map $\Gamma$ has a large
kernel. Some spin networks, such as the ``hourglass'' (see figure \ref{fi_hour}) are mapped to zero \cite{Rovelli:2004tv}.  

Diffeomorphism invariant quantities can give rise to well defined operators on
$\hdiff$. An example is the total volume $V_\Sigma$ of $\Sigma$. The
corresponding operator on $\hkin$ extends to $\hdiff$, thus one obtains a well
defined notion of quantum volume. Areas of surfaces and volumes of subregions
of $\Sigma$ can similarly be quantized, provided surfaces and regions can be
defined in a diffeomorphism invariant fashion, for example by using a matter
field as reference system. 

\section{The Hamilton constraint}
\label{se_ham}
The most complicated constraint is the Hamilton constraint
\begin{equation}
H=\underbrace{\frac{1}{2}\epsilon^{IJK}\frac{E^a_I E^b_J}{\sqrt{\det
E}}F_{abK}}_{H_E}
-(1+\iota^2)\frac{E^a_I E^b_J}{\sqrt{\det E}}K_{[a}^IK_{b]}^J
\end{equation}
Here we have already denoted by $H_E$ the so-called \emph{Euclidean part} of the
constraint, which we will need later. The quantization of the Hamilton
constraint poses several difficulties. On the one hand, its classical action is
very complicated on the basic fields $A$ and $E$. Therefore methods based on a
geometric interpretation, such as were used to find solutions to the
diffeomorphism constraint, are not available. Its functional form on the other
hand makes it hard to quantize in terms of the basic fields because it contains
(a) the inverse volume element, and (b) the curvature of $A$. (a) is problematic
because large classes of states in $\hilb{H}_\text{diff}$ have zero volume, thus
its inverse tends to be ill defined. There have to be subtle cancellations
between the inverse volume and other parts of the constraint for the whole to be
well defined. 
(b) is problematic, because it curvature can not be quantized in a simple way,
at least on $\hilb{H}_\text{kin}$, due to the nature of the inner product.  
It is thus very remarkable that Thiemann \cite{Thiemann:1996aw,Thiemann:1996av,Thiemann:1997rv,Thiemann:1997ru,Thiemann:1997rt} proposed a family of well
defined Hamiltonian constraints, and partially analyzed the solution spaces. We
can not describe his construction with all details, but we will briefly discuss
the most important ideas. 
\subsection{Thiemann's quantization}
The first ingredient in the quantization is the observation that one can absorb
the inverse volume element in the Hamiltonian constraint into a Poisson bracket
between the connection and the volume:
\begin{equation}
\label{eq_trick}
\epsilon^{IJK}\epsilon_{abc} \frac{E^a_I E^b_J}{\sqrt{\det
E}}=\frac{1}{4\iota}\{A_c^K(x),V_\Sigma\}
\end{equation}
Here $V_\Sigma$ is the volume of the spatial slice. The Poisson bracket can be
quantized as a commutator
\begin{equation}
\{\ldots,\ldots\}\longrightarrow\frac{1}{i\hbar}[\ldots, \ldots], 
\end{equation}
and the volume has a well understood quantization as we have discussed before. A
similar trick can also be used to quantize the extrinsic curvature appearing in
the Hamilton constraint. Thiemann found that 
\begin{equation}
K^I_aE_I^a(x)=\{H_E(x),V_{\Sigma}\},\qquad 
\end{equation}
which can be used to quantize the full constraint, once the Euclidean part $H_E$
has been quantized. 

The second important idea is that solutions to all the constraints must, in
particular, be invariant under spatial diffeomorphisms. Thus it is possible to
define operators for curvature as limit of holonomy around shrinking loops.
While such limits are ill defined when acting on kinematical states, they can be
well defined on states in $\hilb{H}_\text{diff}$. Indeed, Thiemann is able to
give a regulated definition for the constraints, which is such that when
evaluated on states in  $\hilb{H}_\text{diff}$, becomes independent of the
regulator, once it is small enough. 
The Poisson bracket involving $A$ can be approximated as 
\begin{equation}
\label{eq_conn}
\epsilon\vec{e}^a\{A_a(x), V_\Sigma\}\approx \{\int_e A,V_\Sigma \}\approx
-h_e^{-1}\{h_e, V_\Sigma\},
\end{equation}
where $e$ is a curve emanating in $x$, $\vec{e}$ is its tangent in $x$, and
$\epsilon$ its coordinate lenght.  
Curvature is treated as in lattice gauge theory
\begin{equation}
\label{eq_curv}
\epsilon^2 F_{ab}(x)\text{d}\sigma^{ab}\approx\int_{S}F\approx h_{\partial
S}-\one, 
\end{equation} 
where $\text{d}\sigma$ is the area element of the surface $S$ in $x$ and
$\epsilon^2$ its coordinate area. To use the formulas \eqref{eq_conn},
\eqref{eq_curv}, one needs to chose curves and surfaces. In for the Hamilton
constraint, these are made to depend on the graph that the state acted on is
based on, and they are assigned in a diffeomorphism covariant fashion. This
still leaves large ambiguities when the operators acts on states in $\hkin$, but
most of them go away, when acting on $\hdiff$: Only their diffeomorphism
invariant properties matter.  

\begin{figure}%
\centerline{\includegraphics[width=8cm]{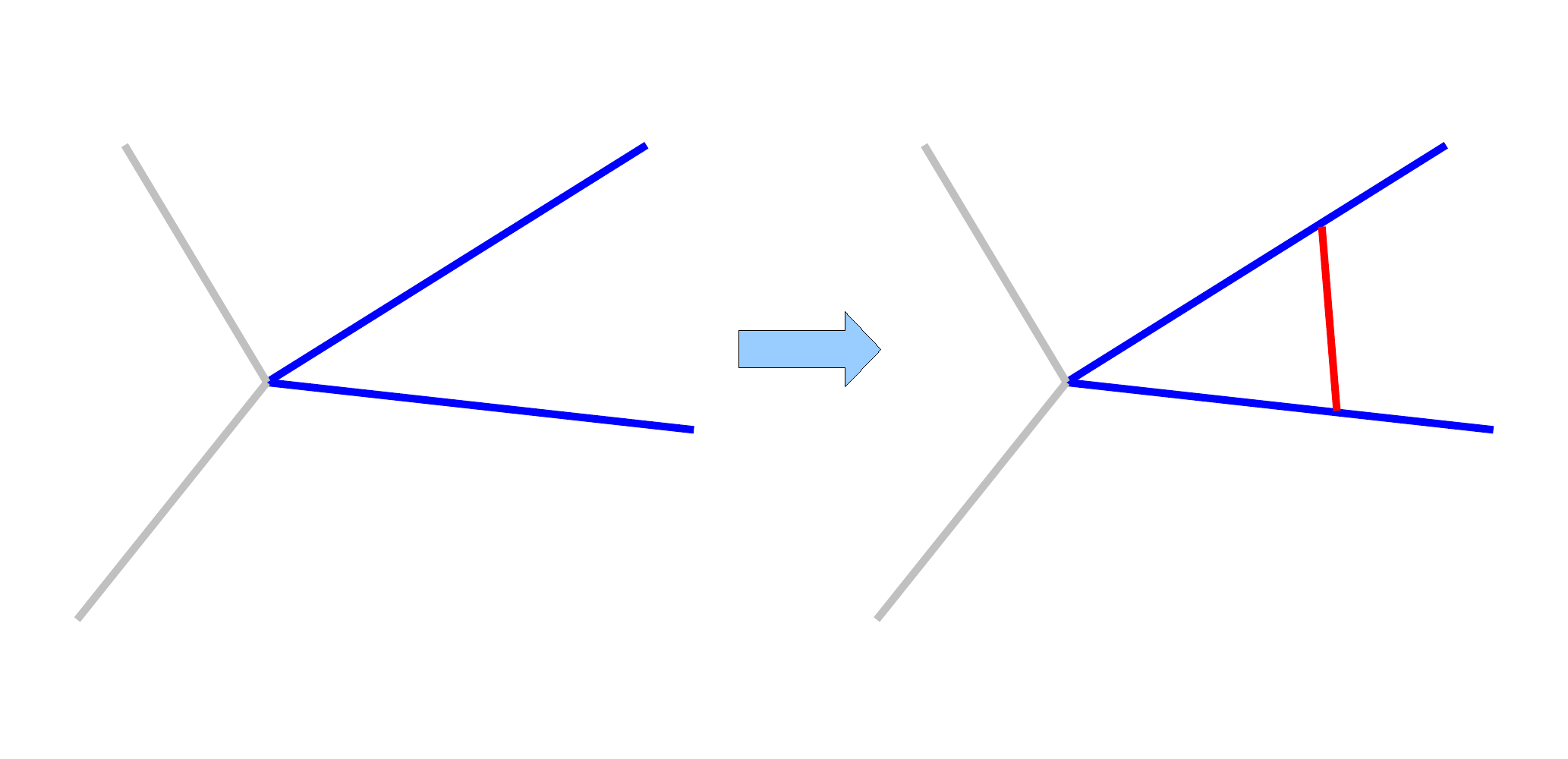}}%
\caption{The Hamilton operator creates new edges between edges incident in a common vertex}%
\label{fi_newedge}%
\end{figure}

Using these ideas, one can define Hamilton constraint operators on $\hkin$, and
by duality on $\hdiff$. The operators  have the following properties:
\begin{itemize}
\item  The action of the constraints is local around the vertices:
\begin{equation}
\widehat{H}(N)\psi_\gamma=\sum_{v \in V(\gamma)}N(v)\widehat{H}(v)\psi_\gamma,
\end{equation} 
where $\psi$ is a cylindrical function based on the graph $\gamma$, the sum is
over the vertices of $\gamma$, and $\widehat{H}(v)$ is an operator that acts
only at, and in the immediate vicinity of, the vertex $v$.      
\item They create new edges (see figure \ref{fi_newedge}). This is because of the use of \eqref{eq_curv}: The
surfaces chosen to regulate the curvature are such that one of the edges
bounding them are not part of the graph of the state acted on.    
	\item They have a nontrivial kernel. 
	\item They are anomaly free in a certain sense: The commutator of two
constraints vanishes on states in $\hdiff$. See the discussion below.   
	\item There are several ambiguities in the definition of the
constraints.
	One is for example the \SUT representation chosen in the regularization
process (in \eqref{eq_conn}, \eqref{eq_curv}, for example, we worked with
holonomies in the defining representation, but other irreducible representations
could be used as well). But there are also ambiguities pertaining to the
creation of new edges, and ambiguities in the application of equations like
\eqref{eq_trick} that are harder to parametrize.   
\end{itemize}
We do not know whether the quantization proposed by Thiemann is the right one.
One important test for the quantization of constraints is whether they
satisfy the relations that are expected from the classical Dirac algebra. The
commutator of two Hamilton constraints is expected to be a diffeomorphism
constraint. Thiemann's quantization is anomaly free in the sense that the
commutator of two Hamilton constraints vanishes when evaluated on a
diffeomorphism invariant state \cite{Thiemann:1996aw}. But it was found that the
same holds when the commutator is evaluated on a much larger set of states, for
which the a diffeomorphism constraint is not expected to vanish
\cite{Lewandowski:1997ba,Gambini:1997bc}. Also there are several ambiguities in
Thiemann's quantization. The meaning of these is largely unclear, but some have
been investigated \cite{Bojowald:2002ny,Perez:2005fn}. Ultimately, the questions
surrounding the quantization of the Hamiltonian should be answered by physical
considerations, for example by checking the classical limit of the theory, or by
other prediction that the theory makes. One situation in which such questions
can be posed and answered is \lqc, and we expect important input from the
findings there. For a technical solution to some of the problems with the
constraint algebra see the next section, on the master constraint program.
Also, substantial progress concerning the dynamics of the theory
has been made in the Spin foam approach, and we hope it will shed light on the
issues, here (see section \ref{se_sfm} and the contribution by Oriti to this
volume). In summary: While many aspects need more study, there is no doubt that
Thiemann's work on the Hamiltoninan contains at least part of the solution of
the problem of dynamics in LQG.
\subsection{The master constraint approach}
\label{se_mc}
As we have pointed out above, the question of whether Thiemann's proposal for
the quantization of the Hamilton constraint is anomaly free is a question that
is not settled. In fact, the Poisson relations between two Hamiltonian
constraints are very complicated and involve the phase space point. The
resulting algebra is thus not a Lie algebra, and it is unclear what a
representation of it should look like. in particular, one expects some quantum
deformations of the structure to occur, but just what constitutes a (harmless)
deformation, and what a (harmful) anomaly is not clear. 
These difficulties prompted the proposal of the master constraint program
\cite{Thiemann:2003zv}. At its core, the proposal is to replace implementation of the infinite
dimensional algebra of constraints with the implementation of one master
constraint. In the case of the Hamiltonian constraint, the proposal is to go
over to the quantity
\begin{equation}
\mathbf{M}=\int_S\,  \frac{(H(x))^2}{\sqrt{\det q}(x)}.  
\end{equation}
It can be argued on general grounds, and checked in examples, that the kernel of
the quantization of such a master constraint  $\mathbf{M}$ is the same as the
joint kernel of the individual constraints constituting the master constraint. 
It is obvious that in this way questions about the constraint algebra can be
alleviated. In the case of \lqg, one can even add squared diffeomorphism and
Gauss constraints to the master constraint above, thus reducing the
considerations to only one constraint altogether. The master constraint is then
much more complicated then the original constraints, but quantization can be
attempted with similar techniques as were used for the Hamilton constraint,
described above. 

The master constraint method has been tested extensively (see for example \cite{Dittrich:2004bp,Dittrich:2004br,Dittrich:2004bs},  and appears to afford
a large simplification in many cases. In eliminating the constraint algebra, it
does however do away with an important check for the correctness of the
quantization. If there are other good ways to check this correctness, this is
no problem, but in cases  -- such as at the present moment the quantization of
the constraints in \lqg -- in which no other good means of checking the
quantization exist, its application is not without danger.
\subsection{Physical inner product, and the link to spin foam models}
\label{se_sfm}
For physical applications it is not merely the physical states that are
important. To compute amplitudes and expectation values one needs an inner
product on these states. In theory, this inner product is obtained from the
constraints themselves. If their joint kernel is contained in the kinematical
Hilbert space, the inner product on that space simply induces one on 
$\hphys$. If zero is in the continuous spectrum of some of the constraints,
there are still mathematical theorems that guarantee the existence of an inner
product, but it can be extremely hard to compute in practice. 

We now want to describe a formal series expansion of the inner product on
$\hphys$ due to Rovelli and Reisenberger \cite{Reisenberger:1996pu}. Since it is
formal, it may not necessarily be useful to calculate the inner product exactly,
but it is hugely important because it makes contact with approaches to quantum
gravity that are starting from discretizations of the path integral of gravity,
so called \emph{spin foam models}. With that, it brings back into \lqg an
intuitive image of time evolution. This is very important even if its physical
merits are still under debate.  

The series expansion is obtained by considering the projector $P_\text{phys}$ on
the Hilbert space $\hphys$ of physical states: Each of the Hamilton constraints
comes with a projector on its kernel. This may be a genuine projection operator,
or a linear map into the dual of $\hdiff$. It can formally be written as 
$P_\text{phys}^{(x)}=\delta(\widehat{H}(x))$.
The projection onto the solution space of the Hamiltonian constraints is the
product of all these projectors. 

\begin{figure}%
\centerline{\includegraphics[width=14cm]{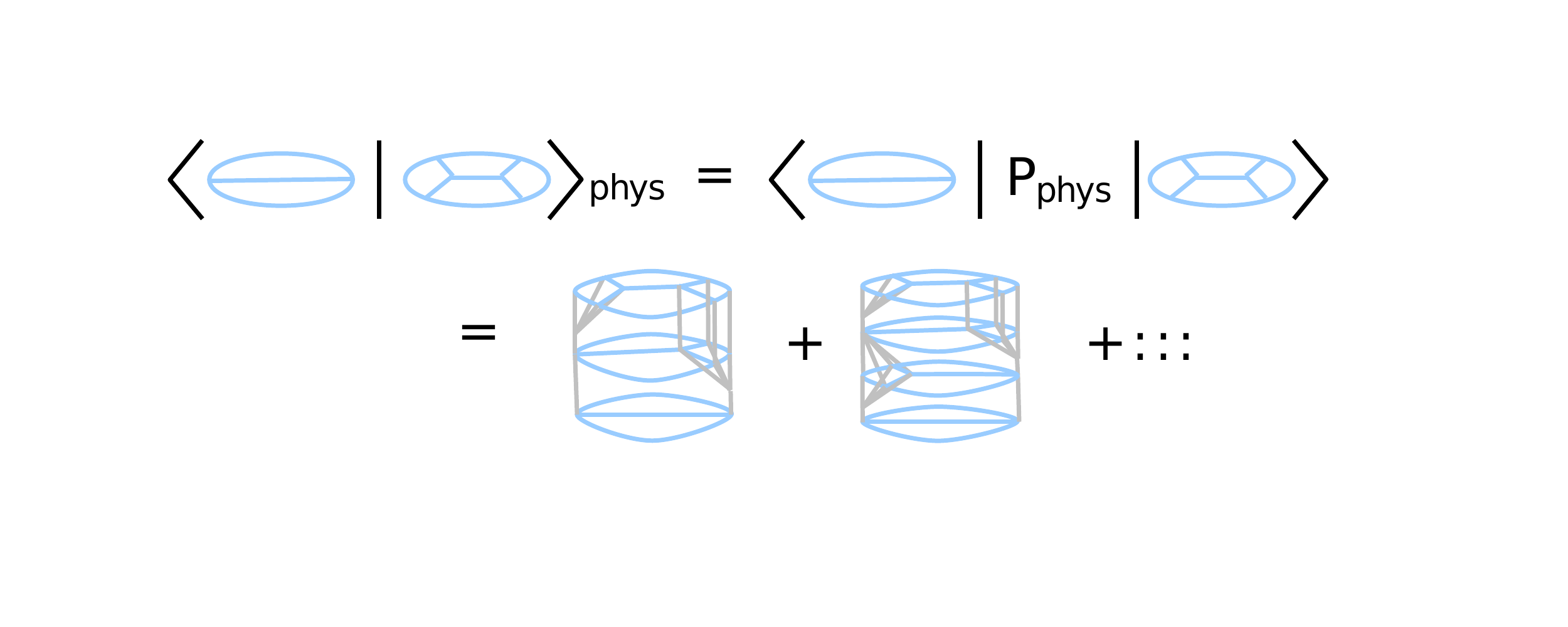}}%
\caption{An illustration of the spin foams occurring in the expansion of the physical inner product of two specific spin networks. Spin labels are not shown for simplicity.}%
\label{fi_spinfoam}%
\end{figure}

The physical inner product between the (physical part of) spin networks $\psi$
and $\psi'$ can be expressed in terms of the projector as
\begin{equation}
\label{eq_physp}
\scpr{P_\text{phys}\psi}{P_\text{phys}\psi'}_{\text{phys}}=\scpr{\psi}{P_\text{
phys}\ldots}_{\hdiff}.
\end{equation}
To obtain the series expansion, one writes the delta functions as functional
integration over the lapse, and expands the exponential:
\begin{equation}
\label{eq_sf}
\begin{split}
P_\text{phys}&=\prod_{x\in\Sigma}\delta(\widehat{H}(x))\\
&= \int DN\, \exp i \int  N(x) \widehat{H}(x)\,\text{d}x\\
&=1+ i\int DN\, \int  N(x) {\widehat{H}}(x)\,\text{d}x\\
&\qquad -\frac{1}{2}\int DN\,\iint  N(x)N(x') { \widehat{H}}(x){
\widehat{H}}(x')\,\text{d}x\,\text{d}x'\\
&\qquad +\ldots.  
\end{split}
\end{equation}
One sees that the expansion parameter is the number of Hamilton constraints in
the expression. It was shown in \cite{Reisenberger:1996pu} how the path
integrals over $N$ can be defined. If one plugs this expansion into
\eqref{eq_physp}, one obtains an expansion of the physical inner product in
terms of the product on $\hdiff$. The Hamilton constraints will create and
destroy edges. In fact, each term in \eqref{eq_sf} will give rise to multiple
terms in the expansion of the inner product, as each of the Hamilton
constraints can, because of the integration over $\Sigma$, act at any of the
vertices. In the end, many of the terms will give zero, however, because the 
scalar product on $\hdiff$ is nonzero only if the graphs underlying the states
are equivalent under diffeomorphisms. This means, that each of the non-zero
terms can be labeled by a diagram that depicts a discrete cobordism, or history,
connecting the two graphs involved in the product. For an illustration, see figure \ref{fi_spinfoam}. In such a diagram, surfaces
represent the evolution of the edges of the spin networks, these meet in lines,
which represent the evolution of the  spin network vertices. 
When a new edge is 
created by the action of a constraint, this is shown in the diagram as a
vertex. Such diagrams, together with the labeling of the surfaces with
representations, and the edges with intertwiners, is called a spin foam. Each 
of these spinfoams is assigned, by the action of the Hamilton constraints and
the inner product on $\hdiff$, a number (amplitude). This is very interesting
for several reasons:
\begin{itemize}
 \item The analogy to Feynman diagrams is striking: In both cases, an evolution
operator is expanded into a series of terms labeled by topological objects with
group representations as labels.
\item Spin foam models have been obtained independently, from discretizations
of the action of general relativity. 
\item Solving the Hamilton constraint means implementing the dynamics of \lqg,
but no notion of evolution is apparent in solutions, at least superficially.
The above expansion brings back a picture of state evolution (although one must
be cautious with simple physical interpretations in terms of geometry
evolving in some specific time).
\end{itemize}
While the connection between \lqg and spin foam models described above is very
convincing in abstract terms,  when one compares the models one gets from using,
for example, Thiemann's constraint, with spin foam models obtained
independently, there are however big technical differences, starting from the
notion of graphs involved (embedded vs.\ abstract), and not ending with the
groups involved. Some of this is changing,  however. For the
interesting new perspectives that result, we refer the reader to \cite{Engle:2007wy,Ding:2009jq,Kaminski:2009fm,Kaminski:2009cc,Alesci:2010gb,Bonzom:2011tf}.

\section{Applications}
\label{se_app}
Now, after the complete formalism of \lqg has been laid out, we can come to
some applications. However, it is presently impossible to solve the constraints
in all generality, and  investigate their physical properties. This is due, on
the one hand, to the difficulties with the implementation of the Hamilton
constraints (see section \ref{se_ham}), and on the other hand to the absence of
useful observables that can be quantized, and used to investigate physical
states. As an example, we remark that the question of whether a space-time
contains black holes or not is well defined, and can in principle be answered in
terms of initial values on a spatial slice $\Sigma$. But to do this in practice
is a very difficult task in the classical theory, and clearly beyond our
abilities in the quantum theory. Therefore, simplifying assumptions, and
approximations have to be made. We will report here on studies on the quantum
theory of a horizon of a black hole (in which the existence of a
null-boundary and some of its symmetries are presupposed), and on some
approximations, called \emph{semiclassical states}, to physical states and
their application to the calculation of matter propagators. Another area with
important physical applications is \lqc, in which the techniques (and in some
cases, results) of \lqg are applied to mini-superspace models. A separate
review is covering this area in detail. Finally we mention the research in spin
foam models which has led to a program to determine the graviton propagator. 
\subsection{Black holes}
\label{se_bh}
Black holes are fascinating objects predicted by general relativity. They even
point beyond the classical theory, because of the singularities within, and
because of the intriguing phenomenon of black hole thermodynamics \cite{Wald:1999vt}.
Therefore they are a tempting subject of investigation in any theory of
quantum gravity. \Lqg was able to successfully describe black
hole horizons in the quantum theory. 
Within this description, it is possible to identify degrees of freedom
that carry the black hole entropy, and prove, for a large class of black holes,
the Bekenstein-Hawking area law. 

The development of this subject is quite rich, with many turns and discussions as to 
the precise definition of the ensemble of quantum states, thus our description will 
leave out many interesting aspects and references. 

The first ideas were developed by Krasnov and Rovelli \cite{Krasnov:2009pd}: Spin network edges pierce the horizon and endow it with area. The number of configurations of these edges (modulo diffeomorphisms) for a given total area is counted to obtain the entropy. 
A systematic and detailed treatment is that by Ashtekar
Baez, and Krasnov \cite{Ashtekar:2000eq} (see also \cite{Ashtekar:1997yu}), in which was realized that the degrees of freedom on the horizon are described by a Chern-Simons theory and are essential in the 
calculation of the entropy. 
\cite{Ashtekar:2000eq} does contain errors in the state counting however, thereby wrongly concluding that only spin network edges with spin 1/2 contribute significantly to the entropy counting. These errors were corrected in 
by Domagala and Lewandowski in \cite{Domagala:2004jt}, where the horizon Hilbert space was correctly derived, its elements characterized in a combinatorial way, and the entropy calculation stated in combinatorial terms and partially carried out. 
It was also shown that the spin 1/2 edges are not generic, and a probability distribution for the edge spins derived. The combinatorial problem was fully solved in \cite{Meissner:2004ju}.
In \cite{Kaul:1998xv,Kaul:2000kf}, Kaul and Majumdar
assumed that a partial gauge fixing that had been used in \cite{Ashtekar:2000eq} was
unnecessary, and they stated and solved the ensuing combinatorial problem for the black hole entropy. They  
thus determined the area-entropy relation in the resulting more natural,
but technically more challenging setting. 
In cite \cite{Engle:2009vc,Engle:2010kt}, it was shown that dropping the partial gauge fixing as in \cite{Kaul:1998xv,Kaul:2000kf} can in fact be fully justified. This led to additional new insights \cite{Bianchi:2010qd}. In our description below, we will follow \cite{Engle:2009vc,Engle:2010kt}.  

There are interesting generalizations (for example \cite{Ashtekar:2004nd,Beetle:2010rd}) and modifications (for example \cite{Ghosh:2006ph,FernandoBarbero:2009ai,Perez:2010pq}) of the formalism. Surprising fine structure has been found \cite{Corichi:2006wn,Corichi:2006bs}
and analyzed \cite{DiazPolo:2007gr,Sahlmann:2007jt,Sahlmann:2007zp,Agullo:2009zt,Agullo:2010zz,G.:2011kb}. 
The later works in this series are remarkable applications of number theory, statistics and combinatorics.
  
The \lqg calculation does not start from solutions of the full theory. Rather,
it quantizes gravity on a manifold with boundary $\Delta$. In the simplest
case, the boundary is assumed to be null, with topology $\R\times S^3$. Again,
there are fields $A$ and $E$ on a manifold $\Sigma$, but now $\Sigma$ has a
boundary $H$. The boundary $\Delta$ is now required to be an \emph{isolated 
horizon}, a quasi-local substitute for an event horizon. This imposes boundary
conditions on the fields  $A$ and $E$ at $H$, 
\begin{equation}
\label{eq_cbound}
 *E=-\frac{a_H}{\pi(1-\iota^2)}F(A). 
\end{equation}
$a_H$ denotes the area of the horizon $H$. 
Furthermore, the symplectic structure acquires a surface term. The latter  
suggests, together with some technical aspects of the kinematical Hilbert space used in \lqg, to quantize the fields on the horizon separately from the bulk fields. The latter are quantized in the way described in section \ref{se_kin}. The only new aspect is that now edges of a spin network can end on the horizon. The such ends of spin network edges are described by quantum numbers $m_p\in\{-j_p,-j_p+1,\ldots, j_p-1,j_p\}$, where $j_p$ is the representation label of the edge ending on the horizon, and $p$ is a label for the endpoint (``puncture''). The quantum number represents the eigenvalue of the component of $E$ normal to the horizon at the puncture.     

The boundary term in the symplectic structure is that of a \SUT Chern-Simons theory with level 
\begin{equation}
k=\frac{a_H}{2\pi\iota(1-\iota^2)l_P^2}, 
\end{equation}
and punctures where spin network edges of the bulk theory end on the surface.
The quantized Chern-Simons connection is flat, locally, but there are degrees of freedom at the punctures. These are -- roughly speaking -- described by quantum numbers $s_p$, $m'_p$, where the former is a half-integer, and $m'_p\in\{-s_p,-s_p+1,\ldots, s_p-1,s_p\}$. There is a constraint on the set of $m'_p$'s coming from the fact that $H$ is a sphere, and hence a loop going around all the punctures is  contractible, and the corresponding holonomy must hence be trivial.   
The Hilbert space is equivalent to a subspace of the singlet component of the tensor product $\pi_{s_1}\otimes \pi_{s_2}\otimes\ldots$ ranging over all punctures. The boundary condition \eqref{eq_cbound} can be quantized to yield an operator equation. The solutions are tensor products of bulk and boundary states in which the quantum numbers $(s_p,m'_p)$ and $(j_p,m_p)$ are equal to each other at each puncture. 

Now, if one fixes the quantum area of the black hole to be $a$, this bounds the number of punctures and the spins $(j_p)$ labeling the representations. It becomes a rather complicated combinatorial problem to determine the number $N(a)$ of quantum states with area $a$ that satisfy the quantum boundary conditions. It was solved in \cite{Kaul:1998xv,Kaul:2000kf}, and later, independently in \cite{Agullo:2009eq}. It turns out that 
\begin{equation}
S(a):=\ln(N(a))=\frac{\iota}{\iota_{\text{SU(2)}}}\frac{a}{4\pi l_P^2}
-\frac{3}{2}\ln\frac{a}{l_P^2}+O(a^0)
\end{equation}
as long as $\iota\leq\sqrt{3}$. Here, $\iota_{\text{SU(2)}}$ is the constant that solves the equation
\begin{equation}
1=\sum_{k=1}^\infty(k+1)\exp\left(-\frac{1}{2}\iota_{\text{SU(2)}}\sqrt{k(k+2)}\right). 
\end{equation} 
One finds $\iota_{\text{SU(2)}}\approx 0.274$. One thus obtains the Bekenstein-Hawking area law upon setting $\iota= \iota_{\text{SU(2)}}$.
\subsection{Semiclassical states and matter propagation}
As we have seen before the trivial spin network is a diff invariant cyclic vector, in a sense, the \emph{vacuum} of \lqg. This state has the spatial geometry be completely degenerate, and the connection field $A$ maximally fluctuating. It is a solution to all the constraints, yet it does not look at all like a classical space-time. Therefore one needs to look for states that behave more like a classical space-time geometry. While it would be desirable 
to find such states that at the same time also satisfy all the constraints, this has not been achieved so far in the full theory (the situation is much better in \lqc, though -- see for example \cite{Ashtekar:2006rx}). Rather, one is settling for states that approximate a given classical metric, and at the same time are approximate solutions to the constraints. Such states have come to be called \emph{semiclassical states}. They are useful for studying the classical limit of the theory, as well as for attempting predictions, and as starting point for perturbation theory.  

One particular class of states that has been studied is using coherent states for the group $\SUT$ \cite{Thiemann:2000bw,Thiemann:2000ca,Thiemann:2000bx,Sahlmann:2001nv}. To understand these states, it is useful to remember the coherent states for the harmonic oscillator: 
\begin{equation}
z:=\frac{1}{\sqrt{2}}\left(\frac{1}{\sigma}{ x_0}+
i\frac{\sigma}{\hbar}{ p_0}\right),\qquad\psi_z^\sigma(x)\sim \left[{
e^{-\sigma^2\Delta}\delta_w}
\right]_{w\rightarrow z}(x)
\end{equation}
Thus coherent states can be viewed as analytic continuations of the heat kernel. This viewpoint makes generalization to a compact Lie Group $G$ possible: 
\begin{equation}
\psi_h^t(g){ :=} \left[\exp\left(-t\Delta^{{ G}}\right)\delta_w^{{G}}
\right]_{w\rightarrow c}(g)\equiv\left[\sum_\pi d_\pi
e^{-t\lambda_\pi}\chi_\pi(gw^{-1})\right]_{w\rightarrow c}(g)
\end{equation}
These states are of minimal uncertainty in a specific sense, and are moreover sharply peaked at a point of $T^*G$ encoded in $h$. These states can be used in \lqg.  The idea is to use a random graph $\gamma$ which is isotropic and homogeneous on large scales,  together with a cell complex dual to 
$\gamma$. In particular there will be a face $S_e$ dual to
each edge $e$ of $\gamma$. Now, given a classical phase space point  
$({A},{E})$ one defines 
\begin{equation}
c_{{ e}}:=\exp\left[i\tau_j\int_{{ S_e}}
    \star { E}^j\right]h_{{ e}}({ A}), 
\end{equation}
and then 
\begin{equation}
\Psi_{\gamma,(A,E)}^t:= \bigotimes_{e\in\gamma}\Psi_{c_e}^t
\qquad\in\cyl_{\gamma}\subset L^2(\overline{\mathcal{A}},d\mu_{AL}).
\end{equation}
These states satisfy the Hamilton constraint weakly, in the sense that the expectation value of the constraint vanishes and they are strongly peaked  at a classical solution.

Such states can be used to approximately compute matter dispersion relations, see for example  \cite{Gambini:1998it,Alfaro:1999wd,Sahlmann:2002qj,Sahlmann:2002qk}. The situation studied is that of a  matter test field propagating on  quantum space-time. Two scenarios have been investigated: 
\begin{enumerate}
	\item The field is coupled to the expectation values (in a semiclassical state) in gravity sector with semiclassical state
	\item The geometry is chosen as  ``typical result'' of a measurement in gravitational sector that has been in a semiclassical state. 
\end{enumerate}
Either case results in a coupling of the matter field to a fluctuating, discrete spatial geometry. In a 1+1 dimensional toy model for a scalar field, 
the dispersion relation has been explicitly calculated \cite{Sahlmann:2009rk}:
\begin{equation}
\omega(k)=c^2 +\ell^2 k^4 +O(k^6)
\end{equation}
with 
\begin{equation}
\begin{split}
c^2&=\lim_{N\rightarrow\infty}\frac{\expec{l}^2}{\expec{l^2}}= \frac{1}{1+\frac{d^2}{l^2}}\\
\ell^2&=\lim_{N\rightarrow\infty}\left(
\frac{1}{N^2}\frac{\expec{l}^4}{\expec{l^2}^3}\sum_{i<j}c_{ij}l_i^2l^2_j-\frac{
N^2}{12} \frac{\expec{l}^4}{\expec{l^2}} \right)\\
&= -\frac{1}{12}l^2\frac{1}{1+\frac{d^2}{l^2}}
\end{split}
\end{equation}
$N$ is here the size of a lattice with periodic boundary conditions, and $\langle\,\cdot\,\rangle$ denotes averages over the random lattice. $l$ is the average effective lattice spacing, and $d$ is a measure of the fluctuation in the latter. The phase velocity $c$ is depending on the details of state and graph, and may do so differently for different fields. This opens the door to  
obtain {severe} constraints on the theory from experiments (see \cite{Klinkhamer:2008ky} for an example).  

We should however point out that since the semiclassical states used in this context are not strict solutions of the constraints, the results obtained with them are only approximations of poorly controllable quality (see for example \cite{Bojowald:2004bb} for a discussion) and should not be interpreted as firm predictions of the theory. As initially stated, the situation is better in \lqc, where semiclassical states that are physical, are available. As an example, the beautiful recent work \cite{Ashtekar:2009mb} applies the ideas of quantum field theory on quantum space-time of \cite{Sahlmann:2002qj,Sahlmann:2002qk} in the context of \lqc. 
\section{Outlook} 
\label{se_out}
\Lqg is a very unusual quantum field theory, and a promising approach to the unification of the principles of general relativity, and quantum theory. But open problems of great importance remain. We have in mind in particular the following questions:
\begin{itemize}
\item Are there restrictions on the types of matter that can be consistently coupled to gravity in the framework of \lqg?  

\item What role does the Barbero-Immirzi parameter $\iota$ play? Can its value be fixed by considerations other than black hole entropy?  

\item How can we extract physics from solutions of the Hamilton constraint?  

\item How can we obtain \emph{controlled} approximations to the solutions of the dynamics? 
\end{itemize}
Progress has already been made on all these. We think that especially the better understanding of the connection to spin foam models and the great results that that have  been achieved in \lqc will help accelerate this progress in the near future. 

\section*{Acknowledgments} 
These notes are an extended version of my talk given at the workshop
``Foundations of Space and Time -- Reflections on Quantum Gravity'' in honor of
George Ellis at the STIAS in Stellenbosch (South Africa). I thank the organizers
of that workshop for their wonderful hospitality, as well as for the great atmosphere they created during the workshop. 

\bibliographystyle{utphys}
\bibliography{proc2}

\end{document}